\documentclass[twocolumn,aps,prb,floats,floatfix,showpacs,superscriptaddress]{revtex4}
\usepackage{amsmath}
\usepackage{graphicx}
\usepackage{dcolumn}
\usepackage{color}

\usepackage{ulem}

\begin{document}

\title{Phonon spectrum, thermal expansion and heat capacity of UO$_2$ from first-principles}

\author{Y.\ Yun}
\altaffiliation[Present ]{address: Laboratory of Reactor Physics ad Systems Behaviour, Paul Scherrer Institut, CH-5232 Villigen PSI, Switzerland}
\affiliation{Department of Physics and Astronomy, Uppsala
University, Box 516, SE-751 20 Uppsala, Sweden}
\author{D.\ Legut}
\affiliation{Nanotechnology Centre, VSB-Technical University of Ostrava,\\
17.\ listopadu 15, CZ-708 33~Ostrava, Czech Republic}
\affiliation{Atomistic Modeling and Design of Materials, University
of Leoben, Leoben, Austria}
\author{P.\,M.\ Oppeneer }
\affiliation{Department of Physics and Astronomy, Uppsala
University, Box 516, SE-751 20 Uppsala, Sweden}
\date{\today}


\begin{abstract}
We report first-principles calculations of the phonon dispersion spectrum, thermal
expansion, and heat capacity of uranium dioxide. The so-called direct method, based on the quasiharmonic
approximation, is used to calculate the phonon frequencies within a density
functional framework for the electronic structure. The phonon dispersions calculated at the theoretical equilibrium volume agree well with
experimental dispersions. The computed phonon density of states (DOS) compare
reasonably well with measurement data, as do also
the calculated frequencies of the Raman and infrared active modes
including the LO/TO splitting.
 To study the pressure dependence of the phonon
frequencies we calculate phonon dispersions for several lattice constants.
Our computed phonon spectra demonstrate the opening of a gap
 between the optical and acoustic modes induced by pressure.
Taking into account the phonon contribution
to the total free energy of UO$_2$ its  thermal expansion coefficient and heat capacity have been {\it ab initio} computed.
Both quantities are in good agreement with available
experimental data for temperatures up to about 500\,K.

\end{abstract}


\maketitle

\section{Introduction}
Over the last few decades UO$_2$ has been one of most widely studied actinide
oxides due to its technological importance as standard fuel material used in
nuclear reactors.
There exists currently considerable interest in understanding
the behavior of nuclear fuel in reactors which is a complex
phenomenon, influenced by a large number of materials' properties, such as
thermomechanical strength, chemical stability, microstructure, and defects.
Especially, knowledge of the fuel's thermodynamic properties, such as specific
heat, thermal expansion, and thermal conductivity, is essential to evaluate the
fuel's performance in nuclear
reactors.\cite{Ronchi07HiTe,Fink00JNM,Carbajo01JNM,Kang06IJT,yunMRS11} These
thermodynamic quantities are directly related to the lattice dynamics of the fuel material.\cite{Bouchet07JAC,Yamada00JAC,watanabe08}

Dolling {\it et al.} \cite{Dolling65CJP} were the first to
measure  phonon dispersion curves of UO$_2$, using the inelastic neutron scattering technique in 1965; their seminal article has become the standard reference for uranium dioxide's
phonon spectrum. Later the vibrational properties of UO$_2$ were investigated in detail
by Schoenes,\cite{Schoenes80} using infrared and Raman spectroscopic techniques. A good
agreement with  phonon frequencies obtained from inelastic neutron scattering was observed.
\cite{Schoenes80}
More recently, Livneh and Sterer \cite{Livneh06} studied the influence of pressure on the Raman scattering in UO$_2$ and Livneh \cite{Livneh08JPCM}
demonstrated the resonant coupling between longitudinal optical (LO) phonons and U$^{4+}$
crystal field excitations in a Raman spectroscopic investigation.
A theoretical investigation  of the phonon spectra of UO$_2$ was reported recently by Yin and Savrasov \cite{Yin08PRL} who employed a combination  of a density-functional-theory (DFT) based technique and a many-body approach. According to their results, the low  thermal conductivity of UO$_2$ stems from the large anharmonicity of the LO modes resulting in no contribution from these modes in the heat transfer.
Goel {\it et al.} \cite{goel07,Goel08JNM} investigated the phonon properties of UO$_2$ using an empirical interatomic potential based on the shell model and observed that the calculated thermodynamic properties including the specific heat are in good agreement with available experimental data.
Devey \cite{Devey11} employed recently the generalized gradient approximation with additional Coulomb $U$ (GGA+$U$) to compute the main phonon mode frequencies at the Brillouin zone center which were in reasonable agreement with experimental data.
Very recently, Sanati {\it et al.}\cite{sanati11} used the GGA and GGA+$U$ approaches to investigate  phonon density of states and elastic and thermal constants, which were found to be in reasonably good agreement with experimental data.
In spite of the already performed studies, further investigations are needed. Especially, the  full dispersions of the phonons in reciprocal space have not yet been considered. Also, important quantities such as the thermal expansion coefficient and heat capacity are directly related to the lattice vibrations but these quantities have not yet been studied {\it ab initio} from the calculated phonon spectrum.

 The objective of this study is to contribute to a detailed understanding of the
lattice vibrations of UO$_2$. Using the first-principles approach, based on the DFT we have calculated phonon dispersion curves and phonon density of states of UO$_2$.
The calculated phonon properties are compared with the available experimental data from inelastic neutron scattering and Raman spectroscopy along with a detailed discussion. Furthermore, several thermodynamic properties have been computed taking the influence of lattice vibrations into account. Here, we report the lattice contribution to the heat capacity as function of temperature as well as temperature and volume (in the quasiharmonic approximation). The dependence of the total free energy on the lattice constant of UO$_2$ as a function of temperature has calculated, from which we derive the  thermal expansion coefficient.
The thermal expansion coefficient as well as  lattice heat capacity compare favorably to available  experimental data up to 500 K, which is the temperature range in which the influence of anharmonicity can be neglected.

\section{Computational Methodology}

The electronic structure of UO$_2$ has been discussed in the past years.
\cite{SDudarev97Phil,crocombette01,RLaskowski04PRB,KKudin02PRL,prodan07,roy08,dorado09,dorado10,petit10,Yun11}
DFT calculations within the generalized gradient approximation (GGA) underestimate the influence of the strong on-site Coulomb repulsion between the $5f$ electrons.
An improved $5f$ electronic structure description can be obtained with the GGA+$U$ approach, in which a supplementary on-site Coulomb repulsion term is added; this approach correctly gives the electronic band gap of UO$_2$.\cite{SDudarev97Phil,prodan07,SDudarev98PhysStat,Yun05NET,Yun07KPS}
While the GGA+$U$ approach would appear preferable for description of UO$_2$'s electronic structure, we encountered specific problems when using this method. Some of the phonon branches became negative away from the zone center. This artifact might be related to the occupation matrix of $5f$ states that would require an additional stabilizing constraint in the GGA+$U$ method.\cite{dorado09,dorado10}
Using conversely the spin-polarized GGA approach, we found that such difficulties did not occur.
The phonon dispersion spectrum presented below is hence computed with the GGA exchange-correlation for antiferromagnetically ordered UO$_2$  and is found to be in good agreement with experiment.\cite{Dolling65CJP}


Here, we have determined the phonon dispersion curves and density of states (DOS) in the quasiharmonic approximation using the direct method.
\cite{Parlinski00PRB,Parlinski03code}  By displacing one atom in a supercell (of 96 atoms) from
its equilibrium position, non-vanishing Hellmann-Feynman forces were
generated. Due to the high symmetry of the face-centered cubic (fcc) lattice of UO$_2$,
only one atom for
uranium (U) and for oxygen (O) was needed to be displaced. The actual shift of the atoms
in the supercell had an amplitude of 0.03 {\AA} and was taken along the [001] direction only, on account of the cubic symmetry of UO$_2$.
In the calculation of the resulting forces we employed the projector augmented wave (PAW) pseudopotential approach within  the Vienna Ab-initio Simulation
Package (VASP).\cite{GKresse93PRB,GKresse96PRB}
The PHONON code
\cite{Parlinski00PRB,Parlinski03code} has been used to extract the force
constant matrix from the Hellmann-Feynman forces and to subsequently
calculate the phonon dispersion curves and DOS.


For the thermodynamic quantities we consider the total free energy of UO$_2$, including the phonon contribution,
\begin{eqnarray}
F(\epsilon,T) = U(\epsilon) + F^{phon}(\epsilon,T) +
F^{el}(\epsilon,T),
\label{eqn:Helmholtz}
\end{eqnarray}
where $F(\epsilon,T)$ is the Helmholtz free energy at a given strain $\epsilon$. The
 phonon free energy contribution $F^{phon}$ is expressed as
\begin{eqnarray}
 F^{phon}(\epsilon,T) &=& \int^{\infty}_0
d{\omega}\,g({\omega},{\epsilon})\big[ {\hbar}{\omega}/2
\nonumber \\
& &+k{_B}T\,
{\rm ln}(1-e^{-{\hbar}{\omega}/k{_B}T}) \big],
\label{eqn:PhononEnergy}
\end{eqnarray}
where  $g({\omega},{\epsilon})$ is the phonon DOS, computed as mentioned above.
 We note that the free electronic energy, $F^{el}(\epsilon,T)$, is not considered in the present study, because the thermal electronic contribution is known to be negligible in the temperature range up to 1000\,K, which is the range of interest in this work.\cite{Sindzingre88,Zhang10}
The static lattice energy $U(\epsilon$) appearing in
Eq.~(\ref{eqn:Helmholtz})  can be expressed as
\begin{eqnarray}
U(\epsilon) = U_0 + V\sum_{ij} C_{ij}{\epsilon}_i {\epsilon}_j,
\label{eqn:StaticEnergy}
\end{eqnarray}
where $U_0$ is the static lattice energy at zero strain, $C_{ij}$ are the
elastic constants, and $V$ is the equilibrium volume at $T = 0$~K. The static lattice
energies have also been calculated using the VASP
code.\cite{GKresse93PRB,GKresse96PRB}

In our calculations we have used a 2$\times$2$\times$2
supercell containing 96 atoms with a 4$\times$4$\times$4 $k$-point mesh in the
Brillouin zone (BZ).
The Perdew-Wang parametrization\cite{perdew92} of the GGA functional was used.
The kinetic energy cut-off for the plane waves was set at 600 eV
and the energy criterion used for convergence was 10$^{-7}$ eV.
The
force acting on each ion was converged until less than 0.01 eV/{\AA}.
Once the phonon DOS has been
calculated, the thermal expansion of UO$_2$ can be evaluated straightforwardly.
First, the phonon DOS with static lattice energy is calculated for several
volumes around the $T=0$~K equilibrium volume. Subsequently, the total free energies are
calculated for these different volumes at constant temperature using
Eqs.~(\ref{eqn:Helmholtz})-(\ref{eqn:StaticEnergy}). After the free energy has been
calculated its minimum gives the corresponding equilibrium volume at the considered temperature.
By repeating the process for different temperatures,
 the thermal expansion coefficient $\alpha$
defined by
\begin{eqnarray}
\alpha (T) \equiv \frac{1}{a}\frac{da}{dT}= \frac{1}{3V}\frac{dV}{dT},
\label{eqn:alpha}
\end{eqnarray}
is obtained; here $a$ is the lattice constant.

A further thermodynamic quantity, the lattice contribution to the specific heat can be derived from
 $\frac{\delta F(V,T)}{\delta T}$ at a fixed temperature in the quasiharmonic approximation.

\section{Results and Discussion}

\subsection{Ground-state properties of UO$_2$}

UO$_2$ crystallizes in the cubic fluorite structure (CaF$_2$) belonging to the $Fm{\overline 3}m$
space group (no.\ 225) and there are three atoms per primitive unit cell with one
U atom (Wyckoff position $4a$) and two inequivalent O atoms (Wyckoff positions
$8c$). Therefore, there are generally nine phonon branches.
Before turning to the description and analysis of the calculated lattice dynamics let us
briefly consider  the ground-state properties of UO$_2$. As mentioned above the employed DFT framework is that of the spin-polarized GGA. The calculated
equilibrium lattice constant {\it a} and bulk modulus {\it B}, which we have obtained by a Birch-Murnaghan $3^{rd}$ order fit,\cite{Birch47}
 are presented in
Table \ref{bulk}, where these are compared to experimental lattice properties.\cite{benedict82,Hutchings87,Yamada97,idiri04}
Our calculated equilibrium volume as well as the bulk modulus compare reasonably well with the experimental data and with results from molecular dynamics simulations.\cite{Yamada00JAC,watanabe08} Compared to the experimental lattice constant the lattice constant computed here  is 1.2{\%} smaller. This can be attributed to a too strong binding of the $5f$ orbitals which become too much delocalized in the spin-polarized-GGA approach. In the GGA+$U$ approach the $5f$ orbitals are more localized and their contribution to the bonding reduced,
\cite{SDudarev97Phil,dorado09,RLaskowski04PRB} which leads to a theoretical lattice parameter which is larger than the experimental one.\cite{Yun11}

\begin{table}
\caption{Calculated equilibrium lattice constant {\it a} (in {\AA}) and bulk modulus  {\it B} (in GPa) of UO$_2$. Theoretical values obtained in this work are compared to values from molecular dynamics simulations\cite{Yamada00JAC} (MD), as well as to experimental data\cite{benedict82,Hutchings87,Yamada97,idiri04} (Exp.).
}
\begin{ruledtabular}
\begin{tabular}{llc}
          & $a$ ({\AA}) & $B$ (GPa) \\
\hline
This work &  5.406   &  184      \\
MD (300\,K)  &  5.472\cite{Yamada00JAC} &  182\cite{Yamada00JAC} \\
Exp.\ (300\,K) &  5.47\cite{Yamada97,idiri04} & 192,\cite{benedict82}\,198,\cite{Yamada97}\,207\cite{idiri04}\\
\end{tabular}
\end{ruledtabular}
\label{bulk}
\end{table}

\subsection{Phonon spectrum of UO$_2$}

\begin{figure}[!bt]
\includegraphics[width=0.99\linewidth]{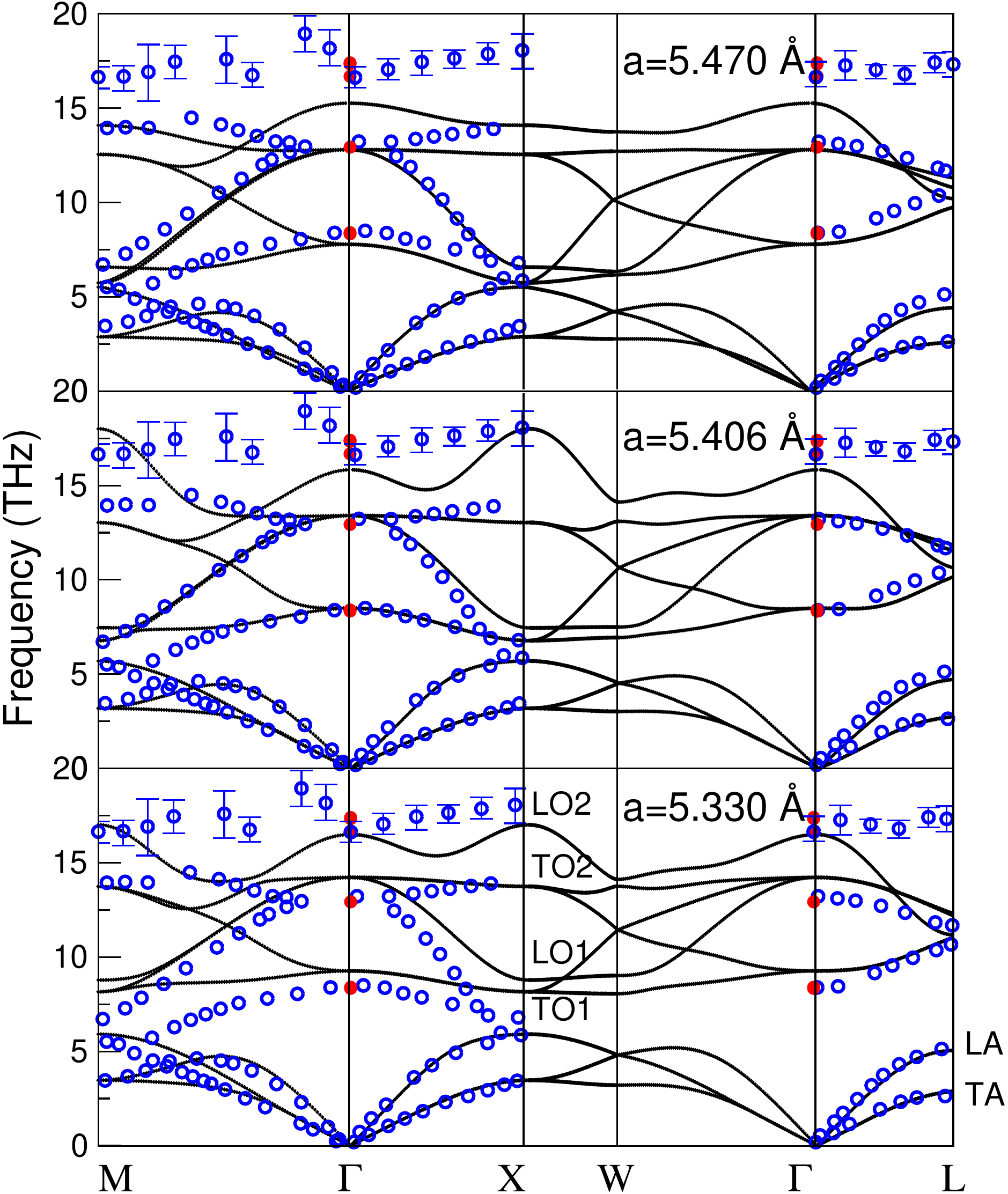}
\caption{(Color online) Calculated (full lines) and measured
 phonon dispersion curves (open symbols, from Ref.\ \onlinecite{Dolling65CJP}) along high-symmetry directions in the fcc
 Brillouin zone.  The measured Raman and infrared modes (Ref.\ \onlinecite{Livneh08JPCM})  at the $\Gamma$ point are depicted by solid (red) circles. The notation of the special points is M: (1,\,1,\,0),
 $\Gamma$: (0,\,0,\,0), X: (1,\,0,\,0), W: $(1,\, \frac{1}{2},\,0)$, and
L: $(\frac{1}{2}, \frac{1}{2}, \frac{1}{2})$.
\label{fig1}}
\end{figure}

Figure~\ref{fig1} shows the measured and {\it ab initio} calculated
phonon dispersion curves of UO$_2$ along high-symmetry points of the fcc BZ.
%
The black lines are the calculated phonon dispersions obtained for the
different lattice parameters, the blue open symbols are the experimentally measured
data from inelastic neutron scattering,\cite{Dolling65CJP} and red closed symbols are data from Raman scattering.\cite{Livneh08JPCM} Note that M denotes the additional point $(1,\,1,\,0)$ that was included by Dolling {\it et al.},\cite{Dolling65CJP} but that is outside of the first BZ.
The calculated phonon dispersions along the high-symmetry lines  X-W-$\Gamma$ have been added to illustrate the whole dispersion of phonon states in the BZ.
The three branches in the
low frequency region are the transverse acoustic (TA) and longitudinal
acoustic (LA) modes that belong to vibrations of the U atom with its relatively
heavy mass. The other branches are optical modes with higher
frequencies that are mainly associated with  lattice vibrations of O
atoms and can be labeled as TO1, LO1 and TO2 and LO2, respectively (see Fig.\ \ref{fig1}),
as there are two inequivalent O atoms in the unit cell.
The three panels in Figure~\ref{fig1} illustrate the volume dependence of the phonon frequencies, which have been calculated at three different lattice constants, the experimental one, $a=5.47$ {\AA}, the optimized theoretical one, $a=5.406$ {\AA},
and $a=5.33$ {\AA}, a selected smaller lattice constant.

In the surrounding of the $\Gamma$ point the agreement between the calculated and measured phonon frequencies is very good.
A first discrepancy between measured and computed dispersions is observed along the M$-\Gamma$ high-symmetry direction. Our calculations predict three acoustic branches whereas in experiment there are two branches. It could be that a splitting of the low-lying TA branch near the M point could not be sufficiently resolved in the experiment.
 This might be related to the inelastic neutron scattering technique, which might be affected by the size of samples and as well as a relatively low neutron flux (see, e.g., Refs. \onlinecite{Lander80JMMM} and \onlinecite{Lander91JMMM}). Also, as mentioned by Dolling {\it et al.} the frequency measurements of the phonons with the neutron technique might be impeded in the zone boundary
regions.
The LO and TO branches agree reasonably well with experiment, in particular for the optimized theoretical lattice parameter. There are some discrepancies in the positions of the branches at the zone boundary X and at the M point. One of the TO1 branches turning up from the $\Gamma$ point to the M point has not been detected in the experiment. Along $\Gamma -$L the agreement is quite good.
The top-most (LO2) branch deviates most between calculation and the experiment. This high-lying branch has the largest experimental uncertainty. Nonetheless, the {\it ab initio} calculated branch has more dispersion than present in the measurement and, except for the zone center, it falls outside of the experimental error bar.

The phonon frequencies calculated at $a=5.406$ {\AA} and $a=5.47$ {\AA} are very
similar to each other and agree reasonably well with the measurement data.  A notable difference between the two sets of dispersion curves appears however in the frequency
gap at the zone boundaries of M and X.
When the lattice constant decreases with pressure from $a=5.47$ {\AA} to $a=5.33$ {\AA} a  pressure-induced phonon softening occurs. The frequency gap between LA and TO1 modes at the zone boundary X point and at the M point is increased as the pressure increases.
At $a=5.47$ {\AA} the LA and TO1 modes almost approach each other, whereas at $a=5.406$ {\AA}, a gap is predicted to exist between the LA and TO1 modes. A small or vanishing gap between these modes is in accordance with the measurement.
We also note that with increased lattice constant the negative slope of the LO1 branch is
remarkably increased  along the $\Gamma -$X symmetry line. This finding suggests that the propagation of the LO1 phonon is significantly restrained as the lattice constant increases.

In Fig.~\ref{fig1} we furthermore depicted by the red solid circles at the $\Gamma$ point the frequencies obtained by Raman and infrared measurements.\cite{Schoenes80,Livneh08JPCM}
%
Using group theory analysis these active Raman and infrared modes can be decomposed into irreducible representations of the (O$_h^5$) point group,  as
1T$_{2g}+2$T$_{1u}$.
 The U atom contributes only to the infrared active mode (T$_{1u}$), whereas the O atom contributes to both, the infrared and Raman mode, T$_{1u}$ and T$_{2g}$, respectively. Both these modes are triply degenerate. The frequencies of these modes are summarized in Table \ref{R+IR}, together with results from MD simulations\cite{Goel08JNM}  and experimental results.\cite{Dolling65CJP,Schoenes80,Livneh08JPCM} To account for the LO/TO splitting  the Lyddane-Sachs-Teller \cite{LST} relation was used on the basis of effective charges. 
These were chosen to be 3 for the uranium atom and $-1.5$ for oxygen. These values are consistent with values used in the literature.\cite{goel07}
The macroscopic electric field in UO$_2$ splits the infrared-active optical modes into TO and LO components. Frequencies of TO modes are calculated in a straightforward manner within the direct method but the LO modes can only be obtained via introduction of a non-analytical term\cite{Pick70} into the dynamical matrix. In general, this term depends on the Born effective charge tensor and the electronic part of the dielectric function (high-frequency dielectric constant).
Details regarding this procedure can be found elsewhere, see e.g.\ Ref.\ \onlinecite{Wdowik07}.

The agreement of the computed Raman optical mode as well as the infrared TO mode with available experimental data is quite good, see Table \ref{R+IR}. The calculated frequency of the infrared LO mode is somewhat smaller than the experimental results, indicating that the effective charge and dielectric constant taken from literature do not fully account for the correct
$\frac{\varepsilon_0}{\varepsilon_\infty}=\frac{\omega^2_{LO}}{\omega^2_{TO}}$. 
 In addition there is a spread of ca. 0.7 THz in the measured values of the LO infrared mode frequencies.\cite{Axe66,Schoenes80}
This is, however, not sufficient to account for the  underestimation of the experimental values by about 2 THz that is observed for $a=5.470$ {\AA} (upper panel of Fig.\ \ref{fig1}). Using the calculated theoretical volume (middle panel of Fig.\ \ref{fig1}) this deviation is reduced by 25\%. A similar error of ca.\ 1.5 THz for the {\it ab initio} calculated infrared frequencies of another insulating compound, CsNiF$_3$, was reported in a recent study.\cite{Leg10}

\begin{table}
\caption{\label{R+IR} The frequencies  of the Raman (T$_{2g}$) and infra-red (T$_{1u}$) vibrational modes of UO$_2$ at the $\Gamma$-point. Theoretical results (this work)  are given for three lattice constants $a$ and compared to frequencies obtained from molecular dynamics simulation\cite{Goel08JNM} (MD) as well as  experimental (exp.) frequencies (given in THz).}
\begin{ruledtabular}
\begin{tabular}{cccccccc}
Mode  & & \multicolumn{3}{c}{this work} & &  {MD}\cite{Goel08JNM}     & exp. \\
\hline
 $a$ (\AA) &&  5.33 &  5.406 & 5.47 && 5.47 & 5.47 \\
\hline
T$_{1u}$(TO) && 9.27  & 8.49  & 7.78  && 7.62 & 8.34\cite{Axe66}, 8.4\cite{Schoenes80}\\
                    & &         &         &         &&         &         8.52\cite{Dolling65CJP} \\
T$_{1u}$(LO) && 16.5  & 15.82 & 15.26 && 17.28 & 16.68\cite{Axe66}, 16.7\cite{Dolling65CJP}\\
 &&& & & & &  17.34\cite{Schoenes80}, 17.4\cite{Livneh08JPCM} \\
T$_{2g}$    & & 14.22 & 13.4  & 12.80 & & 14.04 & 12.93\cite{Axe66}, 13.42 \cite{Dolling65CJP} \\
\end{tabular}
\end{ruledtabular}
\end{table}

\begin{figure}[b]
\includegraphics[width=0.99\linewidth]{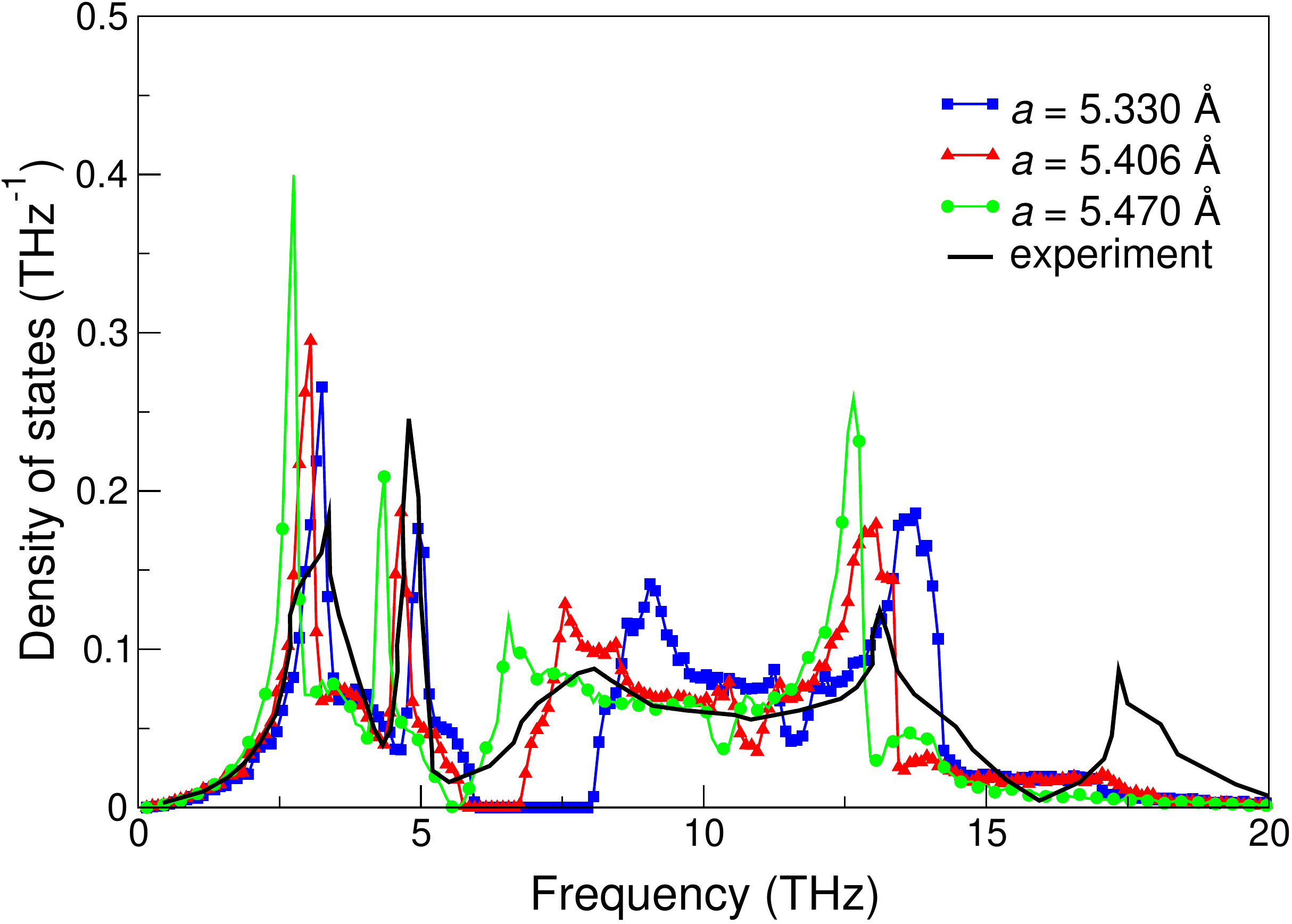}
\caption{(Color online) The theoretical phonon density of states (DOS) of UO$_2$ computed for three lattice constants $a=5.330$, 5.406, and 5.406 {\AA}, compared to the measured\cite{Dolling65CJP} DOS (for $a \approx5.47$ {\AA}, at $T$=296~K). \label{fig2}}
\end{figure}

Figure~\ref{fig2} shows the calculated and measured phonon DOS.
 The blue, red, and green lines with square, triangle, and diamond symbols indicate the phonon DOS computed for $a$=5.330 {\AA}, 5.406 {\AA}, and $5.470$ {\AA}, respectively.  The experimental data\cite{Dolling65CJP} ($a \approx 5.47$ {\AA}, $T$=296\,K) are  plotted with the black line. The U contribution to the calculated phonon DOSs gives rise to a higher intensity and narrower peak widths in the lower frequency region.
The more broadened DOS with lower intensity that occurs in the higher frequency region is mainly derived from the oxygen atoms.
A notable difference between the phonon DOS at the three lattice parameters
is the size and position of the phonon gap occurring for frequencies of about $6-7$ THz.
For the experimental lattice parameter $a=5.47$ {\AA} the computed gap practically closes.
The experimental phonon DOS spectrum at this lattice parameter shows a minimum at about 6 THz, in reasonable agreement, considering some experimental broadening. We note that the trend of decreasing gap with larger lattice constant continues, leading to a closing of the gap
computed for larger lattice constants (not shown).
Overall, the computed phonons DOS of both the theoretical equilibrium ($a=5.406$ {\AA}) and the  experimental lattice parameter are in good accordance with the measured spectrum. The phonon DOS of the theoretical lattice parameter agrees best with the experimental data at higher frequencies (7 to 13 THz), where the peaks coincide with the measured ones. As mentioned earlier, the LO2 mode lies both lower in the computed spectra and is more dispersive than in the measurements.
We note that the recent GGA+$U$ calculations\cite{sanati11} provide a sharper DOS peak at 17 THz, due to a flatter LO2 dispersion near the zone boundaries. At the zone center the LO2 branch lies however much deeper than in the experiment, at 10 THz ({\it vs.} 17 THz in experiment).

\subsection{Thermal expansion of UO$_{2}$}

\begin{figure}[tb]
\includegraphics[width=0.99\linewidth]{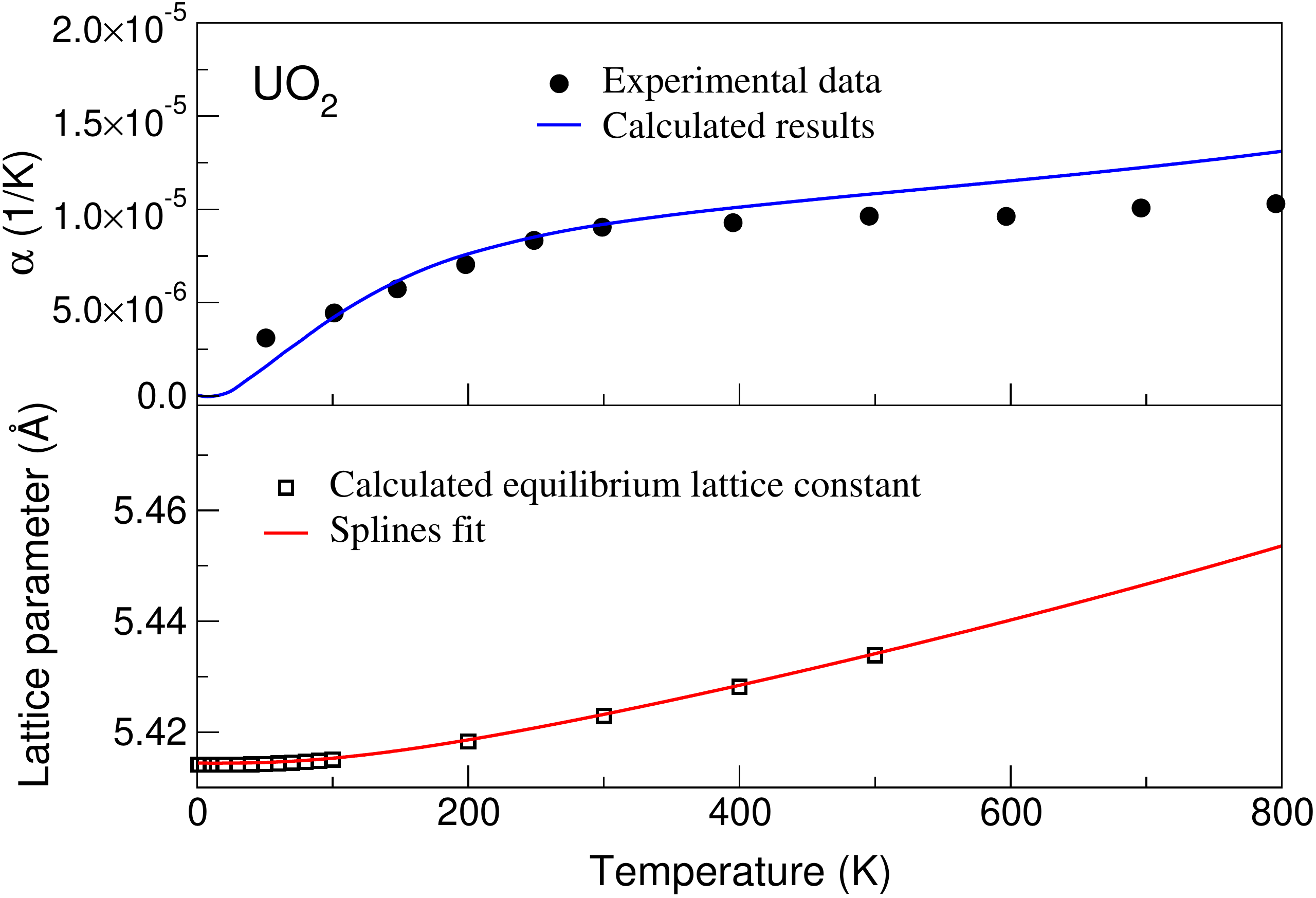}
\caption{(Color online) Top: calculated and experimental thermal expansion coefficient $\alpha (T) $ of UO$_2$. The experimental data are those of Taylor.\cite{Taylor84}
Bottom: Computed temperature-dependent lattice parameter of UO$_2$ (open squares) and spline fit function.
\label{fig3}}
\end{figure}

 The calculated phonon DOS enables us to evaluate some thermodynamic quantities which depend on the lattice vibrations. We start with the thermal expansion.
The phonon contribution to the total free energy
of UO$_2$ increases with increasing temperature and hence becomes progressively responsible
 for changes of the lattice parameters. To compute the thermal expansion of UO$_2$ we have first computed the total free energy, including the phonon contribution, for various lattice parameters, from which we computed the temperature-dependent lattice constant.
Figure~\ref{fig3} (bottom) shows the calculated variation of the lattice constant
with temperature. The red curve gives the spline interpolation of the calculated
lattice constants, shown by the symbols. The thermal expansion coefficient $\alpha (T) =
a^{-1} da/dT$ was subsequently evaluated by differentiating the spline fit.
The upper panel of Fig.~\ref{fig3} shows the calculated thermal expansion coefficient, which is in good agreement with experimental data\cite{Taylor84} up to  500~K. The deviation between the calculated and measured data slightly increases above 500~K and becomes significant at around 1000~K. This might be due to an increased electronic contribution to the thermal expansion.
At very low temperatures, in the region between 0 and 50~K, a deviation is also observed between the calculated and measured data.
The origin of this deviation is not unambiguously clear. We note however that UO$_2$ undergoes a magnetic phase transition at 31~K (see Ref.\ \onlinecite{osborne53}), which may add an additional influence on the lattice parameter.


\subsection{Thermodynamic properties of UO$_{2}$}

\begin{figure}[!tb]
\includegraphics[angle=0,width=0.99\linewidth]{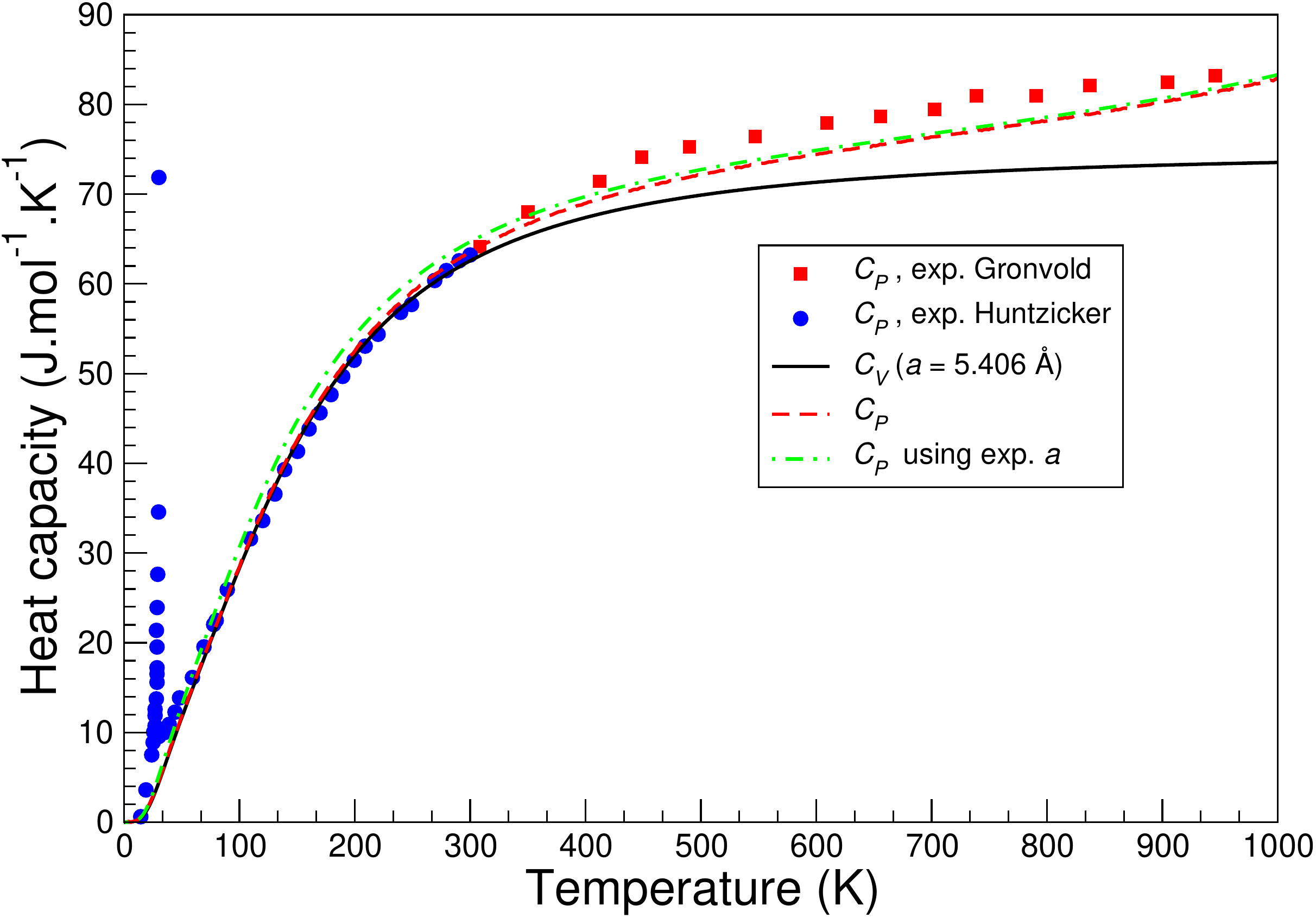}
\caption{
(Color online)  The lattice contribution to the heat capacity of UO$_2$ at constant volume, $C_V$, computed within the harmonic approximation (full curve), and a constant pressure, $C_P$, computed with the quasiharmonic approximation (dashed lines; for $a=5.406$ {\AA} and the experimental $a=5.47$ {\AA}). The experimental data for lower and higher temperatures, full circles and full squares, are taken from Huntzicker and Westrum (Ref.~\onlinecite{huntzicker71}) and Gr{\o}nvold {\it et al.} (Ref.\ \onlinecite{gronvold70}), respectively.
}
\label{hc}
\end{figure}

Next, we employ the Helmholtz free energy to compute the lattice heat capacity at constant volume ($C_V$) and at constant pressure ($C_P$).
Fig.~\ref{hc} shows theoretical results for $C_V$, computed with the harmonic approximation, and $C_P$, computed with the quasiharmonic approximation, as well as experimental results\cite{huntzicker71,fink82} for $C_P$ up to 1000~K.
First the lattice contribution to the specific heat $C_V$ was computed as a function of temperature for the equilibrium lattice constant, $a=5.406$ {\AA}.
Subsequently, the specific heat at constant pressure was derived according to
\begin{equation}
C_P=C_V+9\alpha^2Ba^3T,
\end{equation}
where $\alpha(T)$, $a$, and $B$ are the calculated linear thermal expansion coefficient, the equilibrium lattice constant, and the bulk modulus (see, e.g., Ref.\ \onlinecite{fultz10}).
Fig.~\ref{hc} illustrates that there is a very good agreement between the computed heat capacity
$C_P$ and the experimental data.\cite{huntzicker71,gronvold70,fink82} Note that the sharp anomaly in the experimental data at $T \approx 31$~K is due to the aforementioned magnetic phase transition,\cite{osborne53} which effect is not included in the calculations.
Clearly, the specific heat at constant pressure is in much better agreement with the experimental data at higher temperatures than $C_V$, which is mainly due to the thermal expansion of UO$_2$. Conversely, evaluating $C_P$ for the theoretical equilibrium lattice constant or for the experimental lattice constant $a=5.47$ {\AA} only gives very minor differences.
The computed $C_P$ curves fall somewhat below the experimental $C_P$ data for temperatures in the range of 400 to 1000~K.
The uranium atoms in UO$_2$  are in the paramagnetic state at higher temperatures.\cite{Schoenes80,Hutchings87} Therefore, the remaining difference between experimental and computed data above $T>400$~K could be due to the magnetic entropy contribution to the specific heat or, alternatively, it could be due to the anharmonic effects.
The magnetic entropy contribution to the specific heat was investigated for $T=200$ to 300~K
in Refs.\ \onlinecite{osborne53} and \onlinecite{gronvold70}. At higher temperatures it saturates
to approximately $S = R\, \ln 3$, corroborating a magnetic $J=1$ state on the uranium atoms. It provides to a relatively small magnetic entropy contribution that would lead to a small increase of  the computed $C_P$ data (by about 3 J\,mol$^{-1}$K$^{-1}$).


\section{Conclusions}
We have performed first-principles calculations to investigate the lattice vibrations
and their contribution to thermal properties of UO$_2$. We find that the calculated phonon dispersions are in good agreement with experimental dispersions\cite{Dolling65CJP} measured using inelastic
neutron scattering.
Computing the phonon dispersions for various lattice constants,
we observed a softening of the phonon frequencies with decreasing
lattice constant. Furthermore, the band gap between TA and LO modes at high-symmetry zone-boundary points are found to depend significantly on the volume. This gap almost
closes at $a=5.47$ {\AA}, consistent with a pseudogap detected in the inelastic neutron experiment.
Also, Raman and infrared active modes have been determined as a function of volume. The  agreement with experimental data and with results obtained with molecular dynamics simulations is overall very good, with an exception of the infrared LO mode that appears underestimated in our first-principles calculations.
Including the phonon contribution to the free energy, the heat capacity and the thermal expansion coefficient of UO$_2$ have been computed. Both thermal quantities are found to agree well with experimental data for temperatures up to 500~K.
The good correspondence of the computed and measured thermal data exemplifies the feasibility of performing first-principles modeling of the thermal properties of the important nuclear fuel material UO$_2$.

\begin{acknowledgments}
This work has been supported by Svensk K{\"a}rnbr{\"a}nsle-hantering AB (SKB), the Swedish Research Council (VR), the Swedish National Infrastructure for Computing (SNIC), and by research project No.\ CZ.1.07/2.3.00/20.0074 of the Ministry of Education of the Czech Republic.
D.\ L.\ thanks P.\ Pavone, University of Leoben, Leoben, Austria, and U.\ D.\ Wdowik, Pedagogical University, Cracow, Poland for fruitful discussions.
\end{acknowledgments}

{}
\end{document}